# Irreversibility of field-induced magneto-structural transition in NiCoMnSb shape memory alloy revealed by magnetization, transport and heat capacity studies

**Ajaya K. Nayak[1], K. G. Suresh[1*] and A. K. Nigam[2]**

[1]*Magnetic Materials Laboratory, Department of Physics, Indian Institute of Technology Bombay, Mumbai-400076, India*

[2]*Tata Institute of Fundamental Research, Homi Bhabha Road, Mumbai-400005, India*

The effects of magnetic field on the reverse martensitic transition have been studied in $Ni_{45}Co_5Mn_{38}Sb_{12}$. We find a large field-induced irreversibility in this system, as revealed by the field dependence of resistivity, magnetization and heat capacity data. At the critical temperature, the field-induced conversion of the martensitic to austenite phase is not reversible under any field variation. At this temperature any energy fluctuation induces nucleation and growth of the equilibrium austenite phase at the expense of the metastable martensitic phase and gets arrested. All these three measurements completely rule out the coexistence of austenite and martensitic phases in the irreversibility regime.

*Corresponding author (email: suresh@phy.iitb.ac.in, Phone: +91-22-25723480)

Ferromagnetic shape memory alloys (FSMA) have generated a great deal of interest among the researchers mainly because of their multifunctional properties. Interestingly, many of these alloys also show anomalous magnetic and other related properties arising from the first order nature of the martensitic transition. In this context much attention has been paid to Ni-Mn-X based FSMA due to the observation of magneto-structural transition around the room temperature. This transition can be induced by changing the temperature or the applied magnetic field. Due to the field-induced transition, a large shape memory effect has been observed in NiCoMnIn[1] and $Ni_2MnGa$[2] etc. The martensitic transition is accompanied by a change in the magnetic state, giving rise to large magnetocaloric effect and giant magnetoresistance (MR).[3-8] To realize these properties, a large difference in the magnetization between the austenite and martensitic phases is required, which can be achieved by a proper choice of Mn: X ratio or by some substitution like Co in place of Ni/Mn.[1,9,10]

In many Heusler alloys, the high temperature austenite phase is more ferromagnetic (FM) than the low temperature martensitic phase, which is known to have some antiferromagnetic (AFM) component. Application of a suitable field in the martensitic region causes a field-induced reverse martensitic transition to the austenite phase. This first order transition, in general, leads to the supercooling/superheating of the high/low temperature phase. The magnitude of the field required for this transition depends on the sharpness of the transition and the selected temperature in the transition region. The effect of magnetic field on the martensitic transition has been studied in various compounds.[10-13] In a recent work, the effect of pressure on the martensitic transition and magnetocaloric effect has been reported in NiCoMnSb Heusler alloys.[14] In order to understand the features associated with the martensitic transition on various magnetic and related properties, we have focused on the alloy $Ni_{45}Co_5Mn_{38}Sb_{12}$. A detailed study of the magnetic, magneto-transport and heat capacity properties has been carried out to address the effect of field cycling near the transition region.

The methods of sample preparation and structural characterization have been reported in ref. 9. The magnetization measurements have been performed using a vibrating sample magnetometer attached to a Physical Property Measurement System (Quantum Design, PPMS-6500). The electrical resistivity measurements were carried out using linear four probe method using PPMS. The heat capacity measurement was also performed using PPMS. For the field dependence of resistivity, magnetization and heat capacity measurements, the sample was initially cooled to 200 K and then heated back to the desired measuring temperature.

Figure 1 shows the temperature dependence of heat capacity ($C_P$) measured in zero and 50 kOe fields. Both the measurements were performed in the heating mode after zero field cooling to 2 K. The martensitic transition is accompanied by a large peak in the $C_P$. The temperature of this peak is found to shift from 267 to 259 K as the field is changed from zero to 50 kOe. The transition temperatures are identified as the temperatures where the $C_P$ curve starts changing the slope. In the zero field curve the austenite start ($A_S$) and finish ($A_F$) temperatures are 255 K and 272 K respectively. The inset of figure 1 shows the hysteresis (around the martensitic transition) between the heating and cooling cycles of the thermomagnetic curves measured in 1 kOe. From the cooling curve, the martensitic start ($M_S$) and finish ($M_F$) temperatures are identified as 267K and 245K respectively and from the heating curve the austenite start and finish temperatures ($A_S$ and $A_F$) are noted as 260 K and 280 K respectively. Therefore, there is a small difference between the (austenite) transition temperatures obtained from the magnetization and heat capacity data. This is because the changes in $C_P$ and magnetization occur at slightly different stages in the disorder-broadened first order transition. Similar variations have been reported earlier as well.[15]

The field dependence of resistivity {$\rho(H)$} at different temperatures near the martensitic transition is shown in Fig. 2. The five loop $\rho(H)$ measured at 260 K is shown in Fig. 2(a). In the virgin curve (curve 1), the resistivity starts from 1.4 m$\Omega$.cm in zero field and decreases slowly with field up to 45 kOe, where it suddenly decreases to 0.85 m$\Omega$.cm in a field of 61 kOe due to field-induced martensitic to austenite transition. With further increase in field, the resistivity decreases very slowly in the austenite phase. On decreasing the field (path 2), the reverse transition from austenite to martensitic phase is observed at around 22 kOe, resulting in a large hysteresis, similar to the one observed in the thermomagnetic curve (inset of Fig. 1). The most interesting fact is that the zero field value in path 2 lies below the zero field value of the virgin curve. The $\rho(H)$ curves in the negative field regime are mirror images of the curves in the positive field side. This indicates that the martensitic phase which was transformed to austenite phase with application of field could not fully recover to its initial state by reducing the field to zero. This indicates the field-driven arrest of the austenite phase. The magnetoresistance defined as $\dfrac{\rho(T,H)-\rho(T,0)}{\rho(T,0)}$ is labeled on the right side axis in Fig. 2.

The $\rho(H)$ at 263 K is shown in Fig. 2(b). The resistivity of the virgin curve starts with nearly the same value as that of the 260 K curve but the field-induced transition from the martensitic to the austenite (at 35 kOe) and austenite to martensitic phases (at 6 kOe) are observed in lower fields than that at 260 K. Also, the difference between the zero field resistivity of virgin curve (curve 1) and that in the curve 2 is more than that at 260 K. This implies that a larger amount of the martensitic phase could not regain the initial state in comparison to that at 260 K. The most significant feature is the anomalous behavior observed in the $\rho(H)$ curves taken at 265 K( Fig. 2 (c)). The zero field resistivity starts around the same value as that of the previous two cases and the field-induced transition occurs at around 20 kOe. A field of about 30 kOe has converted a substantial fraction of the material to the austenite phase, as revealed by the insensitiveness of the resistivity on field. On decreasing the field to zero, the resistivity does not change, instead retains a constant value for all further increasing, decreasing or reversing field cycles. This implies that unlike at 260 and 263 K, at 265 K, even a small energy fluctuation could induce the nucleation and growth of the equilibrium austenite phase at the expense of the metastable martensitic phase. This shows that the austenite phase gets permanently locked, resulting in a large irreversibility in the phase transition.

To further investigate the irreversibility observed in the $\rho(H)$ curve, we have measured the field dependence of magnetization {$M(H)$} at 260 K (shown in figure 3) and at 264 K (shown in the inset of Fig. 3). The $M(H)$ loop measured at 260 K shows the metamagnetic transition around 42 kOe in the virgin curve. This indicates the field-induced transition from the low magnetic martensitic state to the high magnetic austenite phase. On decreasing the field, the reverse transition is observed at 16 kOe. In the negative field regime, the observed field-induced transitions are consistent with that of positive field regime. Like in the resistivity case, in this case also, the virgin curve lies outside the envelope curve (5[th] loop). The $M(H)$ loop measured at 264 K shows features similar to that seen in the $\rho(H)$ loop measured at 265 K. The field-induced metamagnetic transition in the virgin curve is observed at about 20 kOe. On decreasing the field, there is no reverse austenite to martensitic transition. The 3$^{rd}$ and 4$^{th}$ loops in the negative field regime do not show any hysteresis and the 5$^{th}$ loop follows exactly the same path as that of the 2$^{nd}$ loop. This confirms that at 264 K, once the martensitic phase is converted to the austenite phase due to the field induced-transition, it cannot regain its previous state by any field change. This irreversibility is seen both in the magnetization and the resistivity data. Similar irreversibility in resistivity and magnetization has been reported in Pr doped maganites.[16] It has been reported that $Nd_7Rh_3$ also shows similar features in the magnetization and MR isotherms due to the supercooling and the kinetic arrest[17]. However, unlike in the present case and the manganites, in $Nd_7Rh_3$, though the resistivity isotherms completely rule out the coexistence of FM and AFM phases in the irreversibility regime, the magnetization isotherms do indicate the presence of coexistence of these two competing phases.

In order to further confirm the irreversibility revealed by the resistivity and the magnetization isotherms, we have measured the dependence of the heat capacity on field cycling in the same temperature regime. The $C_P(H)$ measured at 260 K is shown in Fig. 4(a). The field-induced transition is observed as a sharp peak. In zero field, $C_P$ starts with a low value and increases almost linearly with field up to 42 kOe and then it rises sharply to the peak value at 48 kOe. Above this field, the heat capacity decreases rapidly to less than the value obtained at the initial zero field. In the decreasing field cycle, the $C_P$ increases sharply at 38 kOe and attains the peak value at 33 kOe and then decreases towards the initial zero field value. The hysteresis obtained in $\rho(H)$ and $M(H)$ is also observed in $C_P(H)$. When the temperature is increased to 267 K, the nature of the $C_P(H)$ plot changes considerably. The virgin curve starts from a higher value than that at 260K and decreases with field. The higher value is due to the mixed martensitic and austenite phase around the transition region, which has a larger heat capacity value as observed from the $C_P$ versus $T$ curve (Fig. 1 ). On decreasing the field, $C_P(H)$ retains the high field value of the virgin curve. On all further increasing/ decreasing/reversing field cycles, the heat capacity does not change appreciably, thereby mimicking the $\rho(H)$ behavior at 265 K and the $M(H)$ behavior 264 K. This gives an additional confirmation to the conclusion of field-induced irreversibility and the metastable nature of the martensitic state around the transition temperature, derived from the magnetization and resistivity data. This data clearly show that there is no change in the austenite/martensitic ratio at the end of the virgin curve and.re-establish the arrested nature of the high field phase.[18]

In summary, we have conclusively shown the existence of field-induced irreversibility in the reverse martensitic transition by combining the field dependence of resistivity, magnetization and heat capacity data. This study quite vividly shows that the arrest of the austenite phase becomes very significant at the critical temperature, thereby giving rise to large irreversibility. The off-stoichiometry of the compound must be playing a role in the arrest of the phase at the disorder-broadened, field-induced first order transition. The consistency that we see in the three measurements completely rule out the coexistence of martensitic and austenite phases in the irreversibility regime.

**References:**

[1]R. Kainuma, Y. Imano, W. Ito, Y. Sutou, H. Morito, S. Okamoto, O.Kitakami, K. Oikawa, A. Fujita, T. Kanomata, and K. Ishida, Nature (London) **439**, 957 (2006)

[2]K. Ullakko, J. K. Huang, C. Kantner, and R. C. O'Handley, V. V. Kokorin, Appl. Phys. Lett. **69**, 1966 (1996)

[3]T. Krenke, E. Duman, M. Acet, E. F. Wassermann, X. Moya, L. Mañosa, and A. Planes Nature mater. **4**, 450 (2005)

[4]M. Pasquale, C. P. Sasso, L. H. Lewis, L. Giudici, T. Lograsso, and D. Schlagel, Phys. Rev. B **72**, 094435 (2005)

[5]A. K. Pathak, M. Khan, I. Dubenko, S. Stadler and N. Ali, Appl. Phys. Lett. **90**, 262504 (2007)

[6]Z. D. Han, D. H. Wang, C. L.hang, H. C. Xuan, B. X. Gu, and Y. W. Du, Appl. Phys. Lett. **90**, 042507 (2007)

[7]M. Khan, N. Ali and S. Stadler, J. Appl. Phys. **101**, 053919 (2007)

[8]S. Y. Yu, L. Ma, G. D. Liu, J. L. Chen, Z. X. Cao, G. H. Wu, B. Zhang, X. X. Zhang, Appl. Phys. Lett. **90**, 242501 (2007)

[9]A. K. Nayak, K. G. Suresh and A K Nigam, J. Phys. D: Appl. Phys. **42,** 035009 (2009)

[10]R. Kainuma, W. Ito, R. Y. Umetsu, K. Oikawa, and K. Ishida, Appl. Phys. Lett. **93**, 091906 (2008)

[11]S. Chatterjee, S. Giri, S. Majumdar and S. K. De, Phys. Rev. B **77**, 224440 (2008)

[12]K. Koyama, K. Watanabe, T. Kanomata, R. Kainuma, K. Oikawa, and K. Ishida, Appl. Phys. Lett. **88**, 132505 (2006)

[13]L. Ma, H. W. Zhang, S. Y. Yu, Z. Y. Zhu, J. L. Chen, G. H. Wu, H. Y. Liu, J. P. Qu, and Y. X. Li, Appl. Phys. Lett. **92**, 032509 (2008)

[14] A. K. Nayak, K. G. Suresh and A K Nigam, J. Appl. Phys. **106**, 053901 (2009)

[15] T. Krenke, M. Acet, E. F. Wassermann, X. Moya, L. Mañosa, and A. Planes, Phys. Rev. B **73**, 174413 (2006)

[16] J. Dho and N. H. Hur, Phys. Rev B **67**, 214414 (2003)

[17] K. Sengupta and E. V. Sampathkumaran, Phys. Rev B **73**, 020406 (2006)

[18] K. Kumar, A. K. Pramanik, A. Banerjee, P. Chaddah, S. B. Roy, S. Park, C. L. Zhang, and S.-W. Cheong, Phys. Rev B **73**, 184435 (2006)

**Figure captions:**

FIG. 1. Temperature variation of heat capacity in zero (squares) and 50 kOe (circles) fields recorded during heating. The inset shows the thermal hysteresis between the heating and the cooling cycles in the thermomagnetic data around the martensitic transition in a field of 1 kOe.

FIG. 2. Field dependence of resistivity measured at (a) 260 K, (b) 263 K and (c) 265 K. The magnetoresistance is shown on the right side.

FIG. 3. Magnetization isotherm measured at 260 K. The inset shows the isotherm at 264 K.

FIG. 4. Field dependence of heat capacity measured at (a) 260 K and (b) 267 K.

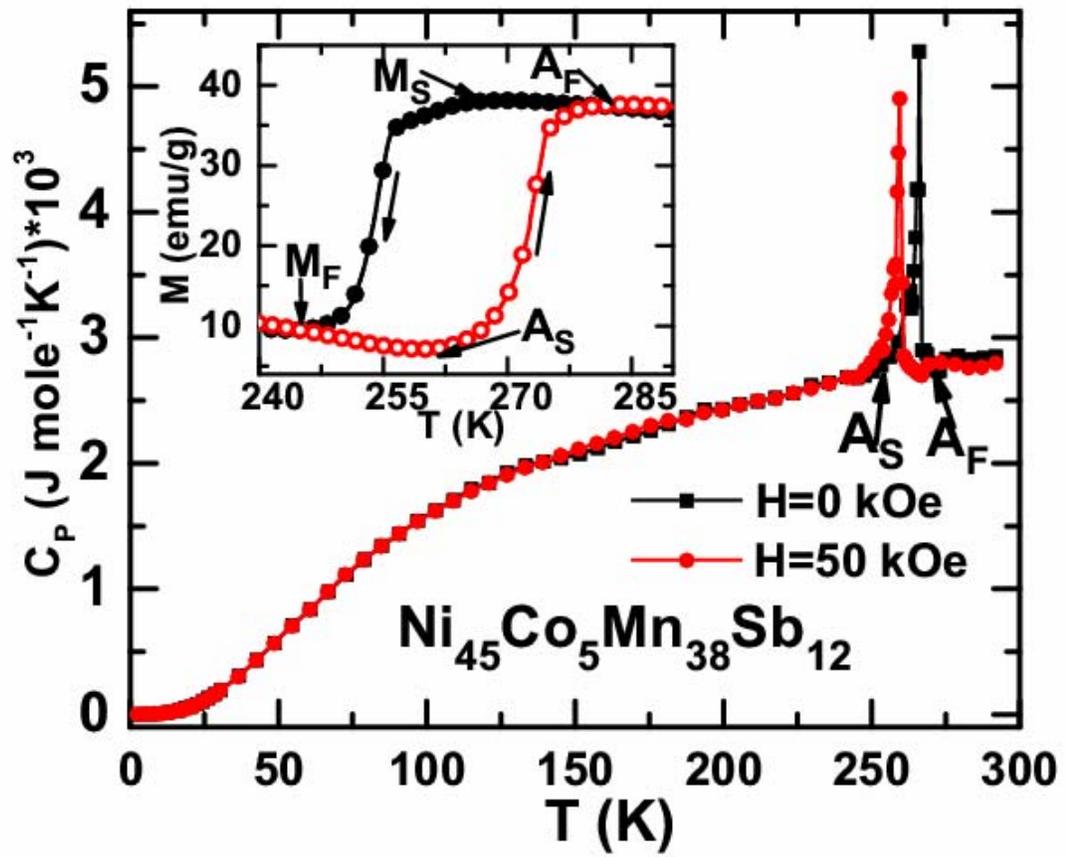
10

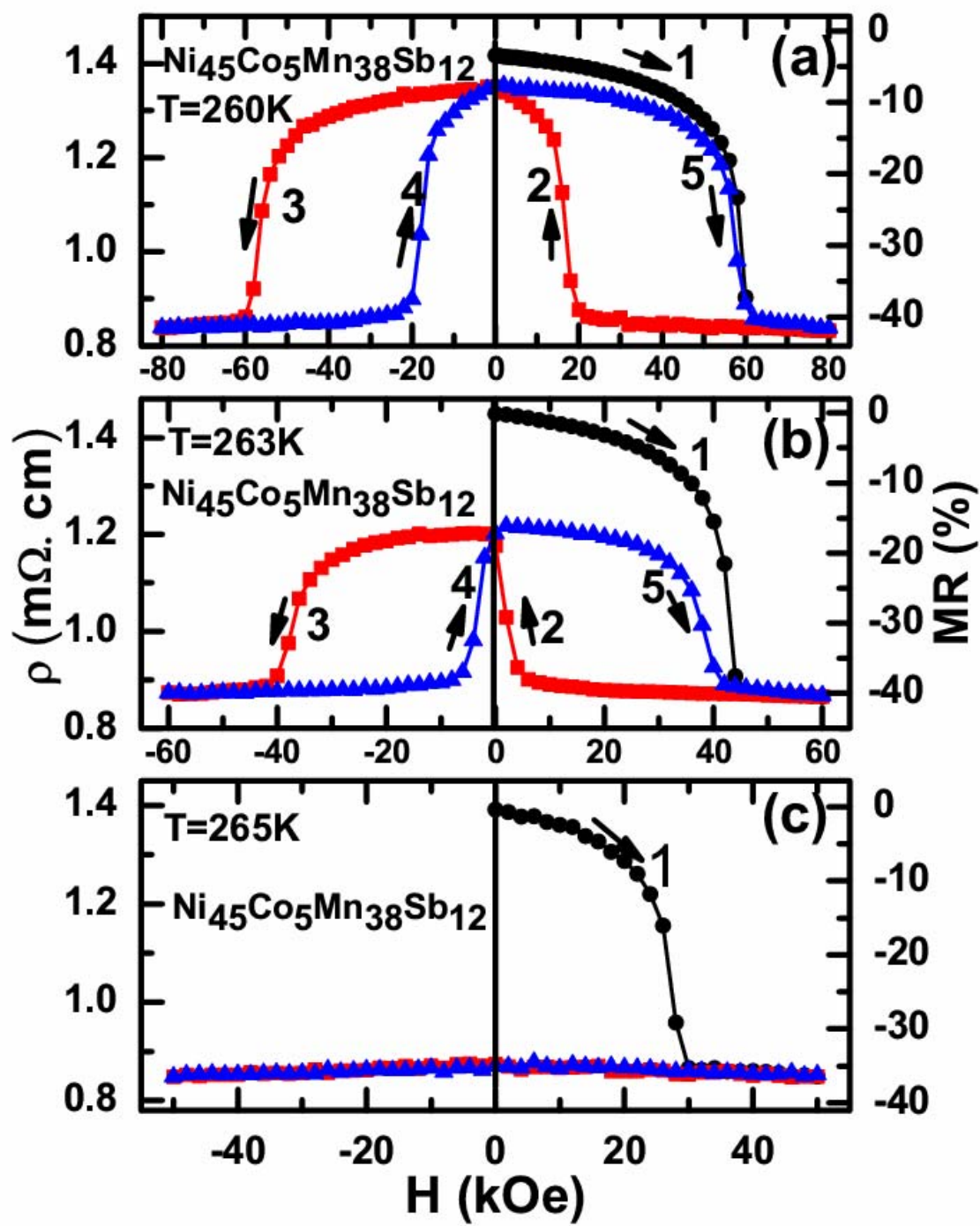

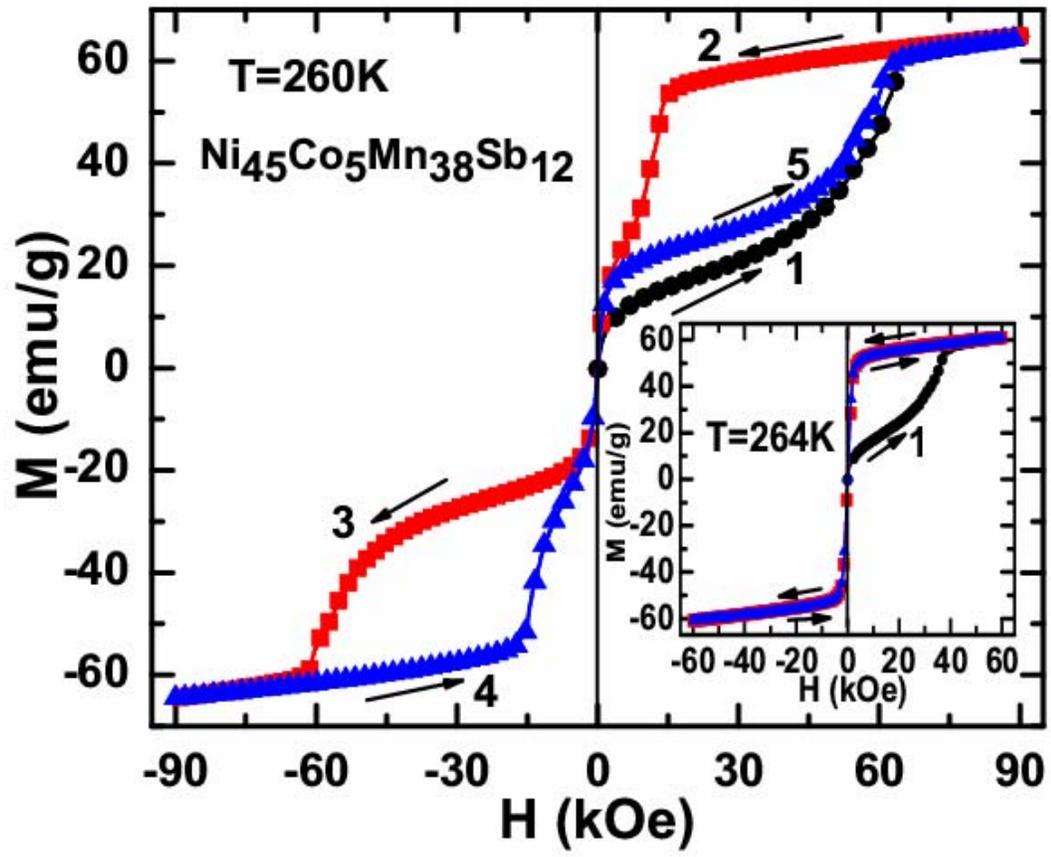

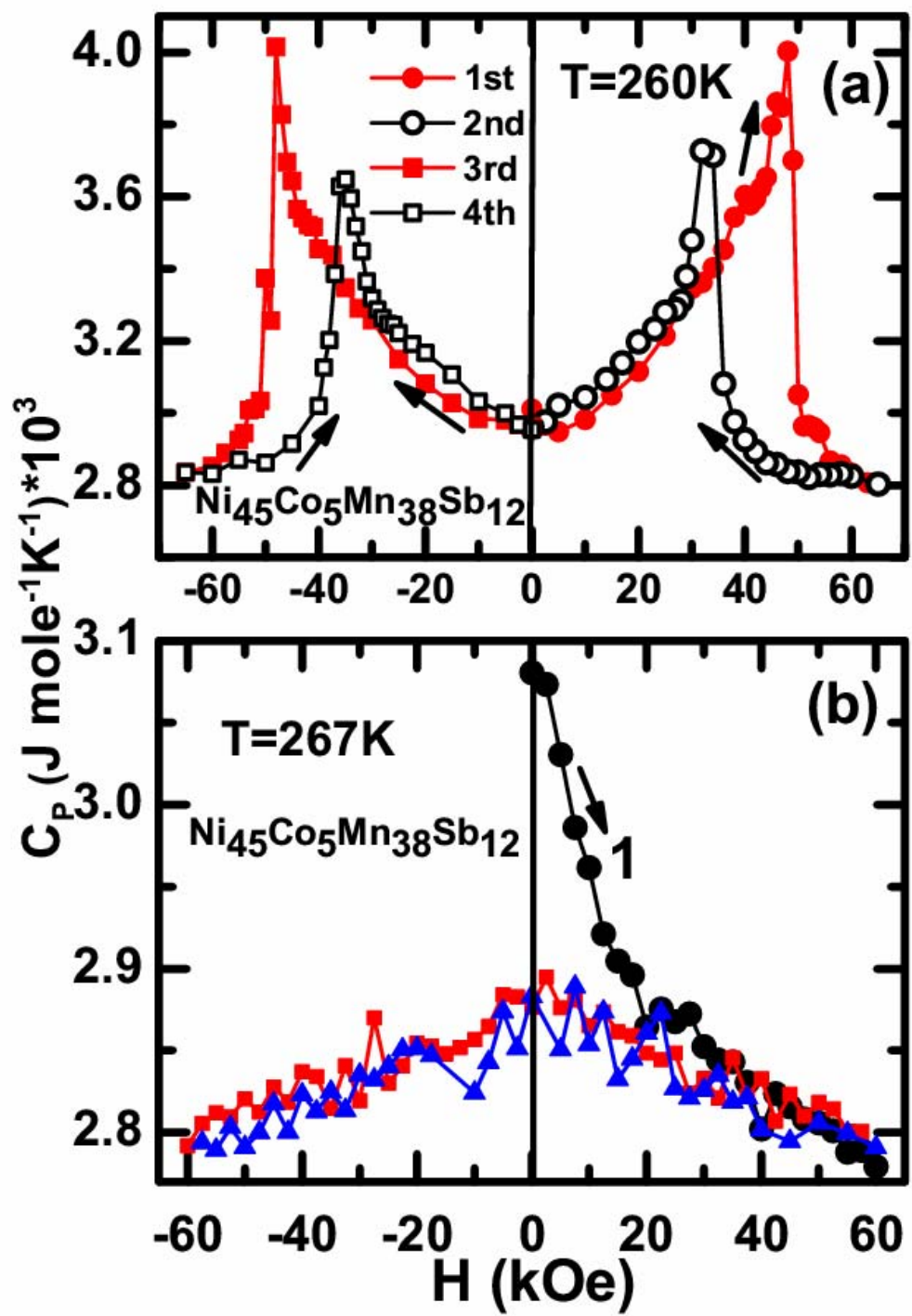